\documentclass[pra,aps,twocolumn,10pt,showpacs,groupedaddress,superscriptaddress,floatfix,notitlepage]{revtex4-2}
\usepackage{amsmath,amsfonts,amssymb,graphics,graphicx,epsfig,color,times, braket}
\usepackage[utf8x]{inputenc}
\usepackage{color}
\usepackage{bbm, dsfont}
\usepackage{subfigure}
\usepackage{mathrsfs}
\usepackage{verbatim} 
\usepackage{xcolor}
\usepackage[colorlinks=true,
linkcolor=blue,
citecolor=blue,
urlcolor=blue,
filecolor=blue]{hyperref}

\begin{document}

\title{High-order quantum correlations in nonlinear waveguide quantum electrodynamics}

\author{I Gusti Ngurah Yudi Handayana}
\email{ngurahyudi@unram.ac.id}
\affiliation{Molecular Science and Technology Program, Taiwan International Graduate Program, Academia Sinica, Taiwan}
\affiliation{Department of Physics, National Central University, Taoyuan City 320317, Taiwan}
\affiliation{Institute of Atomic and Molecular Sciences, Academia Sinica, Taipei 10617, Taiwan}
\affiliation{Department of Physics, Faculty of Mathematics and Natural Sciences, University of Mataram, Indonesia}

\author{Ya-Tang Yu}
\affiliation{Institute of Atomic and Molecular Sciences, Academia Sinica, Taipei 10617, Taiwan}

\author{C.-Y. Lee}
\affiliation{Physics Division, National Center for Theoretical Sciences, Taipei 10617, Taiwan}

\author{K.-T. Lin}
\affiliation{Trapped-Ion Quantum Computing Laboratory, Hon Hai Research Institute, Taipei 11492, Taiwan}

\author{G.-D. Lin}
\affiliation{Physics Division, National Center for Theoretical Sciences, Taipei 10617, Taiwan}
\affiliation{Trapped-Ion Quantum Computing Laboratory, Hon Hai Research Institute, Taipei 11492, Taiwan}
\affiliation{Department of Physics and Center for Quantum Science and Engineering, National Taiwan University, Taipei 10617, Taiwan}

\author{H. H. Jen}
\affiliation{Institute of Atomic and Molecular Sciences, Academia Sinica, Taipei 10617, Taiwan}
\affiliation{Department of Physics, National Central University, Taoyuan City 320317, Taiwan}
\affiliation{Molecular Science and Technology Program, Taiwan International Graduate Program, Academia Sinica, Taiwan}
\affiliation{Physics Division, National Center for Theoretical Sciences, Taipei 10617, Taiwan}

\date{\today}
\renewcommand{\r}{\mathbf{r}}
\newcommand{\f}{\mathbf{f}}
\renewcommand{\k}{\mathbf{k}}
\def\p{\mathbf{p}}
\def\q{\mathbf{q}}
\def\bea{\begin{eqnarray}}
	\def\eea{\end{eqnarray}}
\def\ba{\begin{array}}
	\def\ea{\end{array}}
\def\bdm{\begin{displaymath}}
	\def\edm{\end{displaymath}}
\def\red{\color{red}}
\pacs{}

\begin{abstract}
	Quantum emitters coupled to nonlinear one-dimensional waveguides provide a route for quantum-state engineering by utilizing parametric gain accumulation along with modified waveguide-mediated interactions. Since the accumulated squeezing depends on propagation distance, different emitter separations can experience different effective gain, making the spatial structure of connected quantum correlations a central feature of the dynamics. Here we investigate the transient and steady-state connected correlations of emitter arrays coupled to a parametrically driven waveguide. Using an effective master equation for nonlinear waveguide QED, we analyze second- and third-order connected correlations as functions of the squeezing parameter and interparticle distance. In the two-emitter limit, accumulated squeezing drives excitation buildup and generates a nonzero connected second-order correlation. For many-emitter arrays, tuning the interparticle distance switches the dominant local second-order correlation between the bulk and boundary regions, and enables an analogous spatial control of genuine third-order connected correlations. In the steady state, the averaged second-order correlation exhibits sign-changing and nonmonotonic behavior in the squeezing--interparticle distance parameter space, whereas the third-order connected correlation is strongly enhanced by increasing the squeezing parameter. These results identify nonlinear waveguide QED as a tunable platform for spatially controlling connected quantum correlations and provide insights into correlation engineering in open quantum optical arrays. We further show that this local contrast remains visible in both odd and even number of atomic arrays. 
\end{abstract}
\maketitle

\section{Introduction}\label{sec.introduction}
Waveguide quantum electrodynamics (wQED) provides a controlled framework for coupling quantum emitters through a one-dimensional photonic reservoir, where guided modes mediate long-range coherent and dissipative interactions \cite{Sheremet2023}. This mechanism underlies collective radiation in ordered arrays \cite{Zhang2019,Ke2019,Albrecht2019,Needham2019,Mahmoodian2020,Masson2020,Pennetta2022,Pennetta2022_2}, excitation trapping in dissimilar arrays \cite{Handayana2024}, excitation propagation manipulation \cite{Handayana2026a}, efficient excitation transport in partially driven systems \cite{Yu2025}, and the suppression of high-order quantum correlations in clean--disordered atom--nanophotonic interfaces \cite{Handayana2025}. Recent experimental progress has enabled precise control over emitter--waveguide coupling and propagation phases in nanofiber-coupled atoms \cite{Morrissey2009,Vetsch2010,Solano2017}, photonic-crystal waveguides \cite{Goban2015,Douglas2015}, quantum dots \cite{Luxmoore2013,Arcari2014,Yalla2014,Sollner2015}, superconducting circuits \cite{Roushan2017,Wang2019}, diamond color-center platforms \cite{Riedel2026}, and integrated atom--nanophotonic interfaces \cite{Chang2018}. Although the waveguide is usually treated as a linear reservoir \cite{Jen2025}, optical nonlinearities can make the guided field evolve nontrivially during propagation. 
For example, a classically pumped $\chi^{(2)}$ waveguide can generate squeezed guided fields through parametric processes \cite{Collett1984,Helt2020,Quesada2020}, building on the broader physics of squeezed light and squeezed reservoirs \cite{Walls1983,Gardiner1986,Dalton1999,Murch2013}, and forming a central resource in nonlinear and integrated quantum photonics \cite{Dutt2024,Weiss2026}. This opens a route to correlated few-photon transport \cite{Roy2011, Mahmoodian2018}, experimentally accessible few-photon nonlinear scattering in nanophotonic waveguides \cite{LeJeannic2021}, strongly interacting photons in one-dimensional channels \cite{Hafezi2012}, attractive photonic states in quantum nonlinear media \cite{Firstenberg2013}, programmable nonlinear quantum photonic circuits \cite{Nielsen2025}, and photon-pair generation and statistics engineering in chiral waveguide-QED systems \cite{Tong2026}. 

Recent work has shown that nonlinear wQED can exhibit coherent many-body interactions and decoherence-free dynamics driven by global squeezing fields \cite{Karnieli2025}. While this establishes nonlinear wQED as an active-reservoir platform for many-body control, the connected quantum correlations generated by the same nonlinear system remain less explored. This question arises because accumulated squeezing introduces distance-dependent gain and pairing processes, so the nonlinear reservoir can reshape not only the excitation buildup, but also the pair and higher-order connected excitation across the array. This builds on the range-dependent correlations studied in linear atom--nanophotonic systems \cite{Jen2022}, now with an additional distance-dependent gain from the nonlinear reservoir. As sensitive probes of emergent many-body behavior, including quantum criticality \cite{Osterloh2002,Mishra2018}, Kibble--Zurek dynamics \cite{Polkovnikov2005,Zurek2005,Keesling2019}, and avalanche-like delocalization processes \cite{Leonard2023}, quantum correlations provide a useful way to characterize how the nonlinear reservoir redistributes correlations across different spatial ranges. 

\begin{figure}[t]
	\centering
	\includegraphics[width=0.49\textwidth]{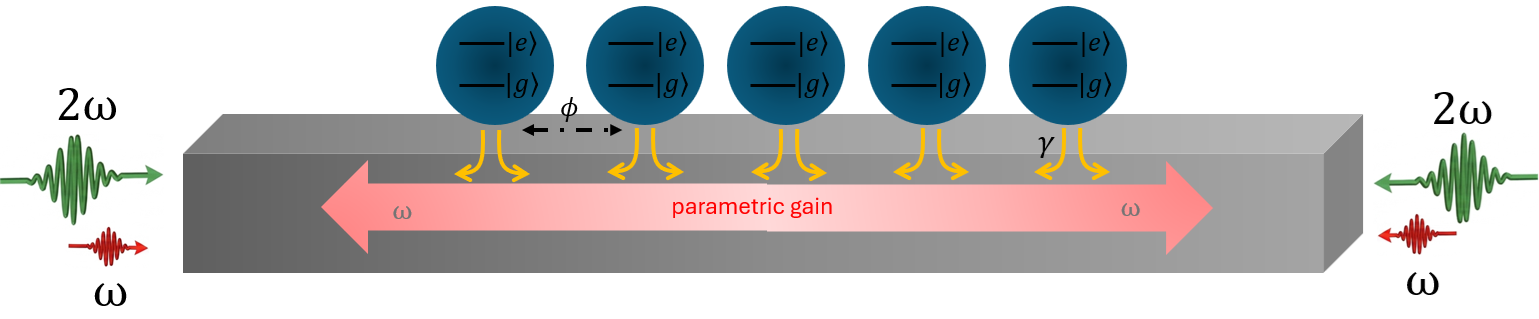}
	\caption{
		Schematic of an array of $N$ two-level emitters coupled to a nonlinear one-dimensional waveguide. 
		Each emitter has ground state $|g\rangle$, excited state $|e\rangle$, and transition frequency $\omega_0$. 
		The emitters couple to the guided-mode continuum, giving rise to the guided decay rate $\gamma$ under the Markov approximation. 
		A classical pump at frequency $\omega_p\simeq 2\omega$ induces broadband parametric gain for guided field fluctuations around the signal frequency $\omega$. 
		The dimensionless interatomic distance is $\phi=k_0\Delta x$, where $k_0$ is the guided-mode wavenumber evaluated at the emitter transition frequency and $\Delta x$ is the nearest-neighbor distance.
	}\label{Fig1}
\end{figure}

The finite geometry of an emitter array provides another route to reveal the spatial structure of nonlinear waveguide-mediated correlations. We focus here on local nearest-neighbor correlations formed near the boundary and in the bulk, comparing pairs with the same emitter separation but different locations in the array. This distinction becomes important in the nonlinear setting because squeezed guided fields accumulate gain during propagation: near-boundary local correlation generated through more asymmetric contributions from the two propagation directions, whereas in the bulk it can be built up from more balanced left- and right-propagating contributions. Finite arrays can also support nonlocal correlations between boundary and bulk emitters, but these correlations involve different emitter separations and accumulated gains, making it difficult to isolate the local boundary--bulk asymmetry from range-dependent nonlinear amplification. 
The mirror-symmetric configuration, schematically shown in Fig.~\ref{Fig1}, makes mirror-symmetric correlations equivalent and therefore provides a convenient way to select representative local correlations near the boundary and in the bulk. Using this local boundary--bulk perspective, we analyze connected second- and third-order correlations as functions of the squeezing parameter and the interparticle distance. We show that changing the interparticle spacing switches the dominant local second-order correlation between bulk- and boundary-dominated regimes. The squeezing parameter produces distinct signatures in different correlation orders: second-order correlations are built up most efficiently at intermediate squeezing, whereas third-order correlations are generated more strongly with increasing squeezing. The resulting local boundary--bulk contrast persists in the steady state and remains visible for either even or odd numbers of atoms in the array.

The paper is organized as follows. In Sec.\ref{sec.theoretical}, we introduce the theoretical model for nonlinear wQED and the effective master equation used throughout this work. In Sec. \ref{sec.two_emitter}, we analyze the two-emitter dynamics as a benchmark and discuss the emergence of second-order connected correlations. In Sec. \ref{sec.many_emitters}, we extend the analysis to many-emitter arrays and identify the transient boundary--bulk competition in second- and third-order correlations. In Sec. \ref{sec.steady_state}, we study the steady-state correlation as a function of squeezing parameter and interparticle distance. In Sec. \ref{sec.finite_size}, we compare odd- and even-emitter arrays and examine their effects on the local boundary--bulk contrast. Finally, we discuss and conclude in Sec. \ref{sec.discuss}.

\section{Theoretical model}\label{sec.theoretical}
We consider an array of $N$ identical two-level emitters coupled to a nonlinear one-dimensional waveguide, as schematically shown in Fig.~\ref{Fig1}. Each emitter has ground state $|g\rangle$, excited state $|e\rangle$, and transition frequency $\omega_0$. The lowering and raising operators of emitter $j$ are denoted by $\sigma_{j}$ and $\sigma^\dagger_{j}$, respectively. The emitters are placed at positions
\bea
x_j=x_1+(j-1)\Delta x,
\qquad
0\le x_1\le x_j\le x_N\le L,
\label{Eq.position}
\eea
where $\Delta x$ is the nearest-neighbor distance and $L$ is the nonlinear waveguide length. The propagation phase accumulated between neighboring emitters is
\bea
\phi=k_0\Delta x,
\label{Eq.phi_def}
\eea
with $k_0$ the guided-mode wavenumber evaluated at the emitter transition frequency. The emitters are assumed to couple uniformly to the guided-mode continuum with guided decay rate $\gamma=2\pi g^2/v_g$, where $g$ is the atom--field coupling constant in the real-space continuum normalization and $v_g$ is the group velocity of the guided mode \cite{Karnieli2025}. This rate sets the natural timescale of the emitter dynamics. We focus on the ideal guided-mode limit and neglect additional nonguided losses.

The waveguide is pumped by a classical field with frequency $\omega_p\simeq 2\omega$, generating broadband parametric gain around the signal frequency $\omega$, as shown in Fig. \ref{Fig1}. In the absence of an external coherent signal drive, the field entering the waveguide at $\omega$ is treated as vacuum guided fluctuations. 
The $\chi^{(2)}$ pump mixes the annihilation and creation components of these fluctuations, thereby producing a squeezed guided field. As a result, the reservoir mediating the emitter--emitter interaction is no longer a passive vacuum channel, but is modified by the accumulated squeezing during propagation.

The reduced dynamics of the emitter array is described by the master equation \cite{Karnieli2025}
\bea
\dot{\rho}
=
-i[H,\rho]
+
\sum_{s=\rightarrow,\leftarrow}
\mathcal{D}_{L_s}[\rho],
\label{Eq.master}
\eea
where
\bea
\mathcal{D}_{L_s}[\rho]
=
L_s\rho L_s^\dagger
-
\frac{1}{2}
\left\{
L_s^\dagger L_s,\rho
\right\}.
\label{Eq.dissipator}
\eea
Here $H$ is the effective Hamiltonian, while $L_{\rightarrow}$ and $L_{\leftarrow}$ are the collective jump operators associated with the right- and left-propagating guided fields, respectively. This effective master equation follows from the nonlinear wQED model \cite{Karnieli2025}, where the field operators at the emitter positions are related to the retarded input fields by a Bogoliubov transformation and using the SLH cascading rule \cite{Combes2017}.

We consider the pure squeezing-accumulation regime, where no externally prepared squeezed reservoir is injected at the input of the emitter array. Thus, the initial squeezing is set to zero, and the nonlinear effect comes only from squeezing accumulated between emitters. We denote the total accumulated squeezing across the array by $r$, so that the squeezing increment between neighboring emitters is
\bea
\Delta r = \frac{r}{N-1}.
\label{Eq.delta_r}
\eea
The right-propagating field accumulates squeezing $\Delta r(j-1)$ before reaching emitter $j$, while the left-propagating field accumulates squeezing $\Delta r(N-j)$.

Following Eqs. \eqref{Eq.master} and \eqref{Eq.dissipator}, the effective Hamiltonian can be written as \cite{Karnieli2025}
\bea
H = H_{\rm ex}+H_{\rm pair}.
\label{Eq.H_split}
\eea
The first contribution is
\bea
H_{\rm ex}
=&&
\frac{\gamma}{2}
\sum_{i,j=1}^{N}
\sin\left(\phi|j-i|\right)
\cosh\left(\Delta r|j-i|\right)
\nonumber\\
&&\times
\left(
\sigma^\dagger_{j}\sigma_{i}
+
\sigma^\dagger_{i}\sigma_{j}
\right),
\label{Eq.Hex}
\eea
which describes excitation-exchange processes dressed by the accumulated squeezing. The second contribution contains squeezing-induced pair-creation and pair-annihilation terms and is given by
	\bea
	H_{\rm pair}
	=&&
	\frac{\gamma}{2}
	\sum_{i,j=1}^{N}
	\sinh\left(\Delta r|j-i|\right)
	\Bigg\{
	\Theta(j-i)
	\Big[
	e^{-i\theta_{\rightarrow}}
	e^{i\phi(i+j)}\nonumber\\
	&&\times \sigma^\dagger_{j}\sigma^\dagger_{i}
	+
	e^{i\theta_{\rightarrow}}
	e^{-i\phi(i+j)}
	\sigma_{i}\sigma_{j}
	\Big]
	+
	\Theta(i-j)
	\Big[
	e^{-i\theta_{\leftarrow}}\nonumber\\
	&&\times e^{-i\phi(i+j)}
	\sigma^\dagger_{j}\sigma^\dagger_{i}
	+
	e^{i\theta_{\leftarrow}}
	e^{i\phi(i+j)}
	\sigma_{i}\sigma_{j}
	\Big]
	\Bigg\}.
	\label{Eq.Hpair}
	\eea 
Here $\Theta(x)$ is the step function that selects the ordered coupling associated with each propagation direction, and $\theta_{\rightarrow}$ and $\theta_{\leftarrow}$ are the squeezing phases of the right- and left-propagating fields. We note that the pairing terms in Eq. (\ref{Eq.Hpair}) are absent in conventional linear wQED \cite{Sheremet2023, Pichler2015}.

The corresponding collective jump operators are
\bea
L_{\rightarrow}
=&&
\sqrt{\gamma}
\sum_{j=1}^{N}
\Big[
e^{-i\phi j}
\cosh\left[\Delta r(j-1)\right]
\sigma_{j}
\nonumber\\
&&
-
i
e^{-i\theta_{\rightarrow}}
e^{i\phi j}
\sinh\left[\Delta r(j-1)\right]
\sigma^\dagger_{j}
\Big],
\label{Eq.Lright}
\eea
and
\bea
L_{\leftarrow}
=&&
\sqrt{\gamma}
\sum_{j=1}^{N}
\Big[
e^{i\phi j}
\cosh\left[\Delta r(N-j)\right]
\sigma_{j}
\nonumber\\
&&
-
i
e^{-i\theta_{\leftarrow}}
e^{-i\phi j}
\sinh\left[\Delta r(N-j)\right]
\sigma^\dagger_{j}
\Big].
\label{Eq.Lleft}
\eea
Eqs. (\ref{Eq.Hex})--(\ref{Eq.Lleft}) show that the nonlinear waveguide dresses the emitter operators by mixing $\sigma$ and $\sigma^\dagger$. 
Therefore, the reservoir mediates not only collective decay and excitation exchange, but also squeezing-induced pairing processes.

A key feature of the present work is that we do not restrict the interparticle distance to the wavelength-distance condition $\phi=2\pi$. Allowing $\phi$ to vary is essential because the interparticle distance controls the interference between excitation-exchange processes and squeezing-induced pairing processes. As shown below, this provides a direct control parameter for switching the dominant local connected correlation between the boundary and bulk regions.

At the same time, we choose the squeezing phases such that the finite array remains mirror symmetric. This mirror-symmetric configuration provides a controlled setting for comparing local boundary and bulk correlations: mirror-related correlation elements become equivalent, while correlations with the same emitter separation but different positions can still be compared. Under reflection about the center of the array,
\bea
j \mapsto \bar{j}=N+1-j,
\label{Eq.mirror_map}
\eea
mirror-symmetric observables should be identical when the effective Hamiltonian and jump operators are invariant under this transformation. 
The exchange part in Eq. (\ref{Eq.Hex}) is automatically invariant because it depends only on $|j-i|$. 
The pairing terms, however, exchange the right- and left-propagating channels under reflection and acquire phase factors depending on $2(N+1)\phi$ (see Appendix \ref{App.mirrorsymetry}). 
To preserve the mirror-symmetry in the Hamiltonian and jump operators, we obtain the required condition,
\bea
\theta_{\rightarrow}-\theta_{\leftarrow}
=
2(N+1)\phi
+
2m\pi,
\qquad
m\in\mathbb{Z}.
\label{Eq.mirror_condition}
\eea
The same condition ensures that the jump operators are mapped into each other up to irrelevant global phases. This mirror symmetry will be used below to identify representative local correlations near the boundary and in the bulk.

\section{Two-Emitter Dynamics and Second-Order Correlations} \label{sec.two_emitter}

We first examine the small-system limit of two emitters coupled to the nonlinear waveguide. This case serves as a useful benchmark for the master equation in Eq. (\ref{Eq.master}), since the full density matrix contains only sixteen components and the relevant observables can be followed directly. Here we initialize the system in the ground state, $\rho(0)=|gg\rangle\langle gg|$. For this initial condition, only a subset of the density-matrix elements is dynamically coupled to the population and correlation observables.

\begin{figure}[h]
	\centering
	\includegraphics[width=0.49\textwidth]{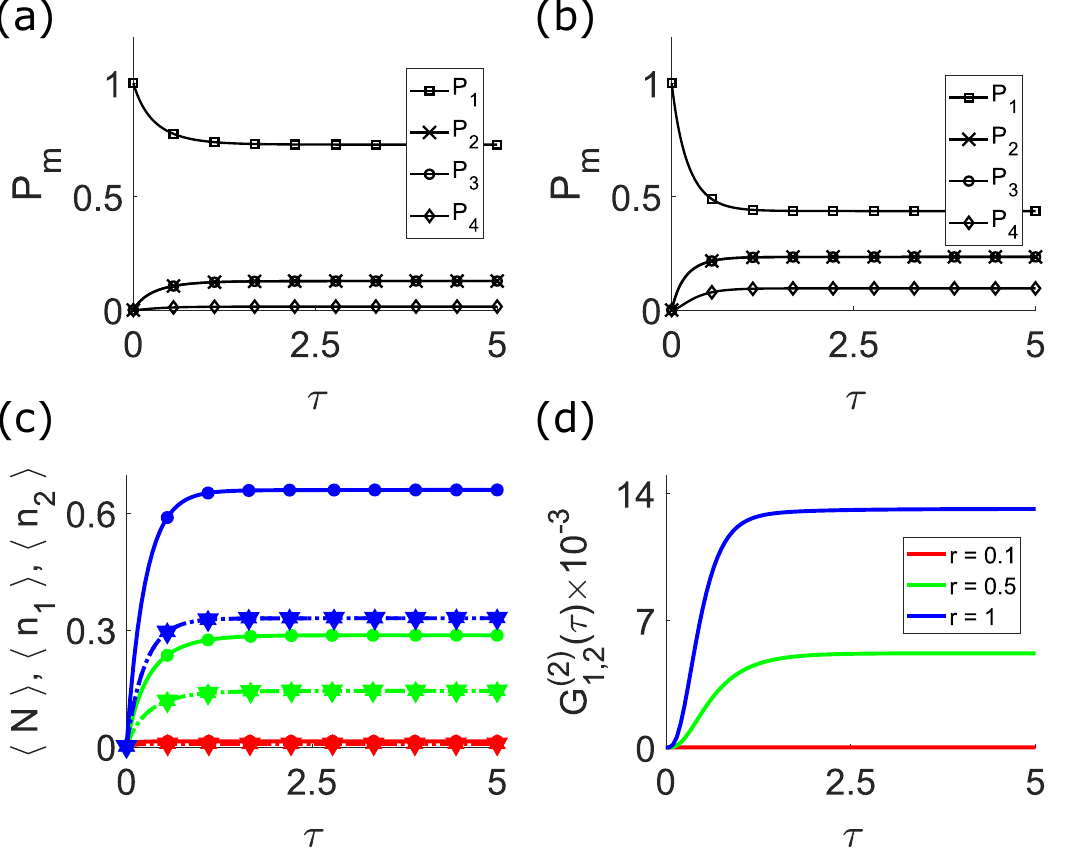}
	\caption{
		Dynamics of a two-emitter system coupled to a nonlinear waveguide.
		(a) and (b) show the state populations $P_m(t)=\langle m|\rho(t)|m\rangle$ for (a) $r=0.5$ and (b) $r=1$, using the ordered basis $\{|m\rangle\} =
		\{|gg\rangle,|eg\rangle,|ge\rangle,|ee\rangle\}$.
		The black curves with square, cross, circle, and diamond markers denote $P_1$, $P_2$, $P_3$, and $P_4$, respectively.
		(c) shows the total excitation number $\langle N\rangle$ and the site populations $\langle n_1\rangle$ and $\langle n_2\rangle$ for $r=0.1$, $0.5$, and $1$, shown in red, green, and blue, respectively.
		For each $r$, solid, dashed, and dotted curves denote $\langle N\rangle$, $\langle n_1\rangle$, and $\langle n_2\rangle$, respectively.
		(d) shows the connected second-order correlation, $G^{(2)}_{1,2}(\tau)$, using the same color convention as in (c).
		Here, $\tau=\gamma r t$ is the rescaled dimensionless time, and the remaining parameters are $\phi=\pi/3$ and $\theta_{\leftarrow}=\pi/2$.
	}\label{Fig2}
\end{figure}
In this calculation, we use the ordered basis
$\{|m\rangle\} =
\{|gg\rangle,|eg\rangle,|ge\rangle,|ee\rangle\}$,
and denote the corresponding state populations by
$P_m(t)=\langle m|\rho(t)|m\rangle$.
Thus, $P_1$, $P_2$, $P_3$, and $P_4$ are the populations of
$|gg\rangle$, $|eg\rangle$, $|ge\rangle$, and $|ee\rangle$, respectively.
We then define the excitation-number operator of emitter $i$ as
$n_i=\sigma_i^\dagger\sigma_i$.
The site populations are the expectation values
$\langle n_i\rangle={\rm Tr}(\rho n_i)$, which in this basis give
$\langle n_1\rangle=P_2+P_4$ and
$\langle n_2\rangle=P_3+P_4$.
The joint excitation probability is obtained from
$\langle n_1 n_2\rangle={\rm Tr}(\rho n_1n_2)=P_4$,
since both emitters are excited only in the state $|ee\rangle$.
The total excitation number is then
$\langle N\rangle=\langle n_1+n_2\rangle=P_2+P_3+2P_4$.
These quantities allow us to monitor how the nonlinear reservoir transfers population from the ground state into the single and double excitation states.
Throughout the dynamical plots, we use the rescaled time $\tau=\gamma r t$, which combines the guided-decay rate $\gamma$ with the accumulated squeezing strength $r$. This scaling is convenient because the accumulated squeezing sets the effective timescale for the nonlinear buildup, while $\gamma$ provides the natural guided-decay rate. 

Figures \ref{Fig2}(a) and \ref{Fig2}(b) show the diagonal density-matrix elements for $r=0.5$ and $r=1$, respectively. In both plots, the ground-state population decreases rapidly at early times, leading to increase in the singly and doubly excited components. A larger value of $r$ leads to larger steady populations in the single- and double-excitation states, reflecting stronger excitation buildup induced by the accumulated squeezing. 

Figure~\ref{Fig2}(c) recasts the population dynamics in Figs.~\ref{Fig2}(a) and \ref{Fig2}(b) in terms of the excitation observables $\langle N\rangle$, $\langle n_1\rangle$, and $\langle n_2\rangle$ for different squeezing parameters. 
The red, green, and blue curves correspond to $r=0.1$, $0.5$, and $1$, respectively, while the solid, dashed, and dotted curves denote $\langle N\rangle$, $\langle n_1\rangle$, and $\langle n_2\rangle$, respectively.
The total excitation number increases with $r$, reflecting the stronger transfer of population from the ground state into the excited-state manifold.
This increase reflects population buildup in both the single- and double-excitation states, with the latter originating from the squeezing-assisted pair-creation terms in Eq.~(\ref{Eq.Hpair}). The site populations $\langle n_1\rangle$ and $\langle n_2\rangle$ overlap, as expected from the mirror symmetry of the two-emitter configuration.

We next quantify whether the two emitters are excited independently or in a correlated manner, quantified by the connected correlation. 
For the two-emitter system, we introduce the connected second-order correlation based on the cumulant expansion \cite{Kubo1962}. 
\bea
 G^{(2)}_{1,2}
 =
 \langle n_1n_2\rangle
 -
 \langle n_1\rangle\langle n_2\rangle.
 \label{Eq.G2_two_def}
\eea
Although $\langle n_1n_2\rangle$ gives the probability that both emitters are excited, this joint probability alone does not determine whether the two excitations are statistically correlated. 
The connected part subtracts the factorized contribution $\langle n_1\rangle\langle n_2\rangle$, which is the joint probability expected from independent site populations. This is also consistent with the quantum-optical viewpoint that correlations are identified through deviations of joint detection or excitation probabilities from factorizable expectations \cite{Glauber1963}.
Using the relations defined above, Eq.~(\ref{Eq.G2_two_def}) becomes
\bea
G^{(2)}_{1,2}
=
P_4
-
(P_2+P_4)(P_3+P_4).
\label{Eq.G2_two_rho}
\eea
 
Figure~\ref{Fig2}(d) shows the time evolution of $G^{(2)}_{1,2}(\tau)$ for the same values of $r$ used in Fig.~\ref{Fig2}(c).
Starting from the ground state, the connected correlation is initially zero and then builds up as the nonlinear reservoir populates the excited states.
For weak squeezing, $r=0.1$, the correlation is finite but remains much smaller than those for $r=0.5$ and $r=1$, so it appears nearly flat on the scale of Fig.~\ref{Fig2}(d).
As $r$ increases, both the buildup and the steady value of $G^{(2)}_{1,2}$ become larger, indicating that accumulated squeezing generates a stronger connected component in the joint excitation probability.
This behavior is consistent with the squeezing-induced pairing terms in Eq.~(\ref{Eq.Hpair}), which couple the ground and doubly excited states through the nonlinear reservoir.
The two-emitter result therefore provides a minimal setting for defining the connected-correlation measure used below in many-emitter arrays.

\section{Transient Higher-Order Correlations in Many-Emitter Array} \label{sec.many_emitters}

We now extend the analysis to many-emitter arrays. 
In contrast to the two-emitter case, the joint quantity $\langle n_i n_j\rangle$ no longer corresponds to the population of a single doubly excited state. Instead, it measures the probability that emitters $i$ and $j$ are simultaneously excited, irrespective of the excitation configuration of the remaining emitters. The connected part therefore isolates the pair contribution beyond independent single-site populations \cite{Kubo1962}. For a general pair, we define
\bea
G^{(2)}_{i,j}
=
\langle n_i n_j\rangle
-
\langle n_i\rangle\langle n_j\rangle .
\label{Eq.G2_many}
\eea
A positive value indicates that the two emitters are excited together more often than expected from independent site populations, whereas a negative value indicates anticorrelation within this connected measure.

Higher-order correlations are introduced within the same cumulant framework. For three emitters, the connected third-order correlation is defined as \cite{Kubo1962}
\bea
G^{(3)}_{i,j,k}
=&&
\langle n_i n_j n_k\rangle
-
\langle n_i n_j\rangle\langle n_k\rangle
-
\langle n_i n_k\rangle\langle n_j\rangle
\nonumber\\
&&
-
\langle n_j n_k\rangle\langle n_i\rangle
+
2\langle n_i\rangle\langle n_j\rangle\langle n_k\rangle .
\label{Eq.G3_many}
\eea
This expression removes all contributions that can be factorized into lower-order moments. Thus, a nonzero $G^{(3)}_{i,j,k}$ means that the simultaneous excitation of sites $i$, $j$, and $k$ cannot be reconstructed solely from the individual populations and pair correlations.

In a finite emitter array, connected-correlation elements with the same separation are not necessarily equivalent, because their locations determine how they are connected to the right- and left-propagating squeezed fields \cite{Karnieli2025}. This is especially relevant in the nonlinear waveguide, where the guided field accumulates squeezing during propagation and the interparticle distance $\phi$ controls the interference between exchange and pairing processes \cite{You2018}. We therefore focus on local nearest-neighbor correlations formed near the boundary and in the bulk. 
This choice keeps the emitter separation fixed while changing the location of the correlated emitters, allowing us to identify a local boundary--bulk contrast without mixing it with longer-range nonlocal correlations.

For $N=5$, mirror symmetry relates the pair $(1,2)$ to $(4,5)$ and the pair $(2,3)$ to $(3,4)$. 
We choose $(1,2)$ as the representative near-boundary nearest-neighbor pair and $(2,3)$ as the representative bulk nearest-neighbor pair, and define
\bea
G^{(2)}_{\rm bd}(\tau)
\equiv
G^{(2)}_{1,2}(\tau),
\qquad
G^{(2)}_{\rm bk}(\tau)
\equiv
G^{(2)}_{2,3}(\tau).
\label{Eq.G2_bd_bk}
\eea
Here, ``bd'' denotes the near-boundary local pair, while ``bk'' denotes the bulk local pair. 
The corresponding local boundary--bulk contrast is
\bea
\Delta G^{(2)}(\tau)
=
G^{(2)}_{\rm bk}(\tau)
-
G^{(2)}_{\rm bd}(\tau).
\label{Eq.DeltaG2}
\eea
A positive value of $\Delta G^{(2)}$ means that the bulk nearest-neighbor pair has the larger connected correlation, while a negative value means that the near-boundary nearest-neighbor pair is larger.

\begin{figure}[t]
	\centering
	\includegraphics[width=0.49\textwidth]{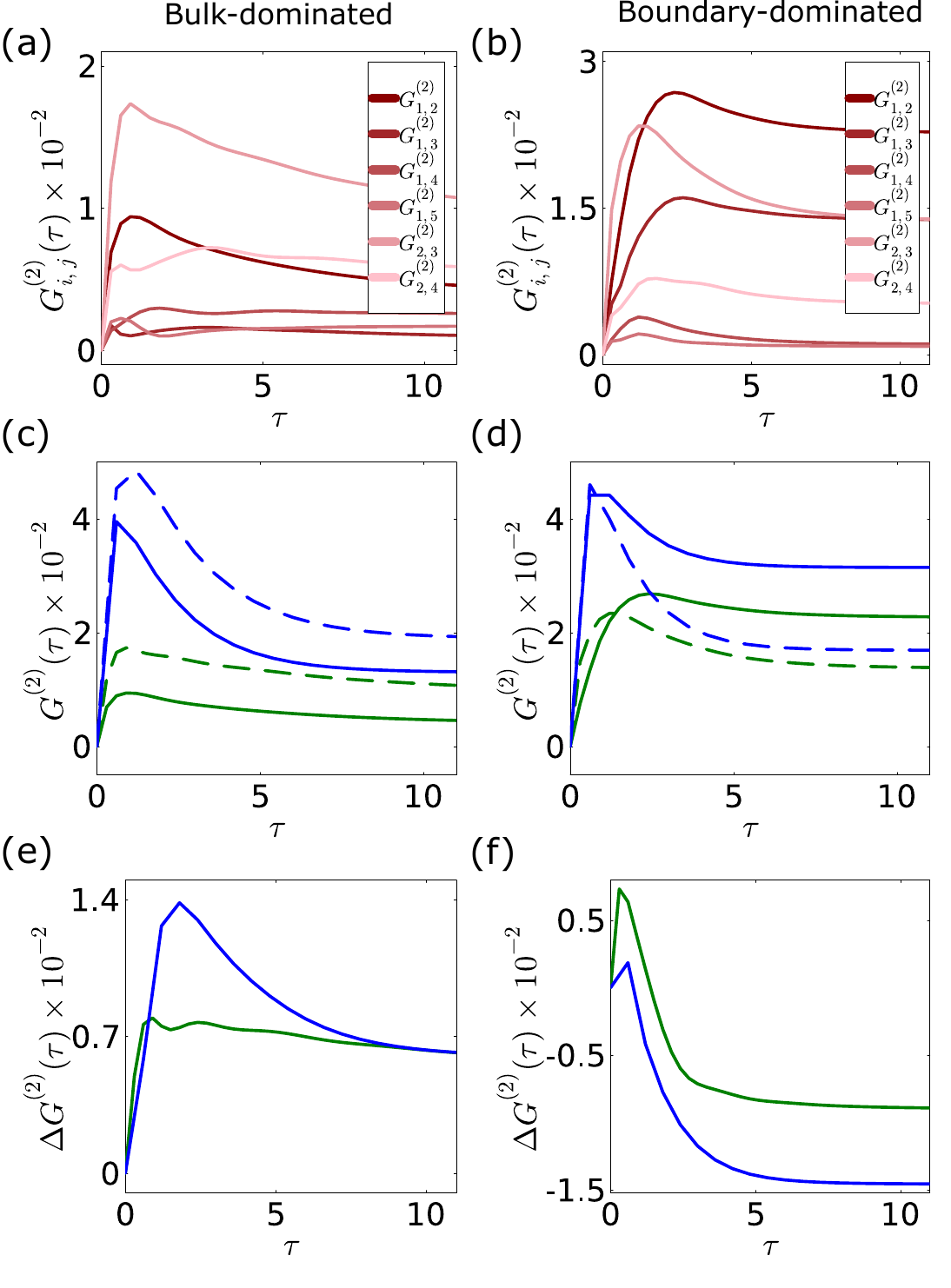}
	\caption{
		Dynamics of the connected second-order correlations for $N=5$. 
		(a) and (b) show the time evolution of the representative pair correlations $G^{(2)}_{i,j}(\tau)$ for $r = 0.5$. 
		The representative pairs are selected according to the mirror symmetry of the array. 
		(c) and (d) focus on the near-boundary and bulk nearest-neighbor correlations, defined as $G^{(2)}_{\mathrm{bd}}(\tau)\equiv G^{(2)}_{1,2}(\tau)$ and $G^{(2)}_{\mathrm{bk}}(\tau)\equiv G^{(2)}_{2,3}(\tau)$, respectively. 
		The solid curves denote $G^{(2)}_{\mathrm{bd}}(\tau)$, while the dashed curves denote $G^{(2)}_{\mathrm{bk}}(\tau)$. 
		(e) and (f) show the corresponding local boundary--bulk contrast, $\Delta G^{(2)}(\tau)=G^{(2)}_{\mathrm{bk}}(\tau)-G^{(2)}_{\mathrm{bd}}(\tau)$. 
		The left column, (a), (c), and (e), corresponds to $\phi=\pi/6$, where the local nearest-neighbor dynamics is bulk-dominated, whereas the right column, (b), (d), and (f), corresponds to $\phi=\pi/4$, where it is boundary-dominated. 
		The line colors in (c)--(f) indicate different squeezing parameters, using the same color convention as in Fig. \ref{Fig2}. 
		The remaining parameter used here is $\theta_{\leftarrow} = \pi/2.$
	}\label{Fig3}
\end{figure}

Figure~\ref{Fig3} shows the transient dynamics of the connected second-order correlations for $N=5$. 
Figures~\ref{Fig3}(a) and \ref{Fig3}(b) show representative pair correlations for $\phi=\pi/6$ and $\phi=\pi/4$, respectively, at $r=0.5$. 
These panels give an overall view of the pair-correlation landscape, including both local nearest-neighbor pairs and nonlocal pairs with larger separations. 
The nonlocal pair correlations can become sizable because they sample longer propagation paths and different accumulated squeezing gains, thereby combining range-dependent amplification with the spatial position of the selected emitters. 
For this reason, the local boundary--bulk comparison below is defined using nearest-neighbor pairs with the same emitter separation, as in Eq.~(\ref{Eq.G2_bd_bk}).

Figures~\ref{Fig3}(c) and \ref{Fig3}(d) compare $G^{(2)}_{\rm bd}$ and $G^{(2)}_{\rm bk}$ for several squeezing parameters. 
The green, and blue curves correspond to $0.5$, and $1$, respectively, following the color convention of Fig.~\ref{Fig2}. 
For $\phi=\pi/6$, the bulk nearest-neighbor correlation is larger than the near-boundary one over most of the evolution, giving the positive contrast shown in Fig.~\ref{Fig3}(e). 
For $\phi=\pi/4$, the trend is reversed: the near-boundary nearest-neighbor correlation becomes larger, leading to the negative contrast in Fig.~\ref{Fig3}(f). 
Thus, the interparticle distance provides direct control over the spatial location of the dominant local second-order correlation, switching it between bulk- and boundary-dominated regimes.
The squeezing parameter controls the correlation buildup, while $\phi$ manipulates the interference between exchange and pairing processes and thereby determines which local correlation dominates. 

\begin{figure}[t]
	\centering
	\includegraphics[width=0.49\textwidth]{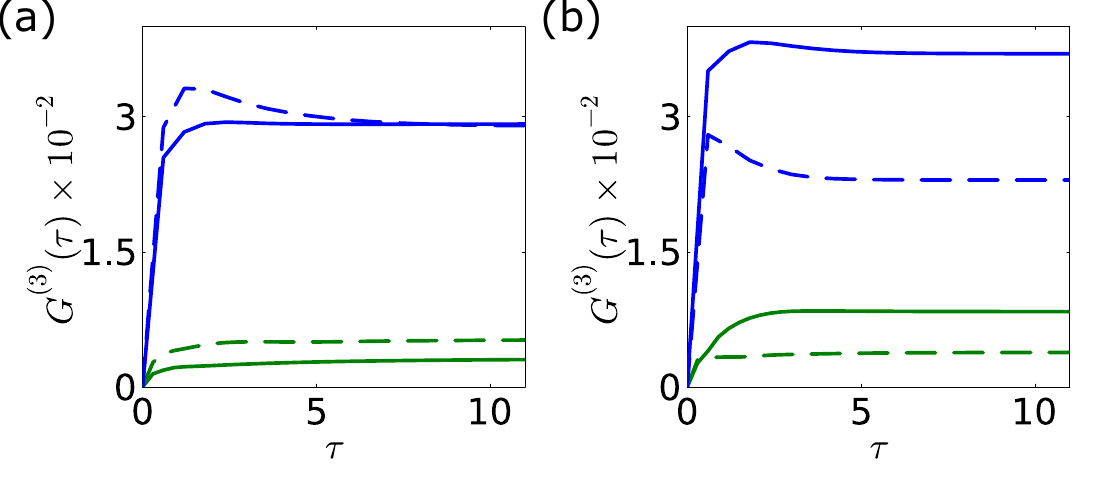}
	\caption{
		Dynamics of the connected third-order correlations. 
		(a) and (b) show the time evolution of the near-boundary and bulk third-order correlations for the same parameter regimes used in Fig.~\ref{Fig3}. 
		The near-boundary and bulk correlations are defined as $G^{(3)}_{\mathrm{bd}}(\tau)\equiv G^{(3)}_{1,2,3}(\tau)$ and $G^{(3)}_{\mathrm{bk}}(\tau)\equiv G^{(3)}_{2,3,4}(\tau)$, respectively. 
		The solid curves denote $G^{(3)}_{\mathrm{bd}}(\tau)$, while the dashed curves denote $G^{(3)}_{\mathrm{bk}}(\tau)$. 
		(a) corresponds to $\phi=\pi/6$, representing the bulk-dominated local regime identified in Fig.~\ref{Fig3}, whereas (b) corresponds to $\phi=\pi/4$, representing the boundary-dominated local regime. 
		The line colors indicate different squeezing parameters, using the same color convention as in Fig. \ref{Fig2}. 
		The remaining parameters are the same as in Fig. \ref{Fig3}.
	}\label{Fig4}
\end{figure}

We next examine whether the same local boundary--bulk contrast appears in higher-order correlations. 
For third-order correlations, we compare the near-boundary local triple $(1,2,3)$ with the bulk local triple $(2,3,4)$,
\bea
G^{(3)}_{\rm bd}(\tau)
\equiv
G^{(3)}_{1,2,3}(\tau),
\qquad
G^{(3)}_{\rm bk}(\tau)
\equiv
G^{(3)}_{2,3,4}(\tau).
\label{Eq.G3_bd_bk}
\eea
Although two atomic positions are overlapped, this minimal consideration shows preliminary explorations of higher-order quantum correlations between the case with one boundary site at $j=1$ and the case all away from the boundary sites. Figure~\ref{Fig4} shows the corresponding third-order dynamics for the same two interparticle distances used in Fig.~\ref{Fig3}. 
For $\phi=\pi/6$, shown in Fig.~\ref{Fig4}(a), the near-boundary and bulk third-order correlations remain close to each other for larger $r$, with the bulk contribution slightly larger in the parameter range shown. 
For $\phi=\pi/4$, shown in Fig.~\ref{Fig4}(b), the near-boundary third-order correlation becomes clearly larger than the bulk one, consistent with the boundary-dominated local second-order contrast in Fig.~\ref{Fig3}. 
The third-order correlations are generated more strongly as $r$ increases, indicating that higher-order connected correlations are highly sensitive to the squeezing-induced buildup of many-emitter excitations.

Together, Figs.~\ref{Fig3} and \ref{Fig4} show that nonlinear waveguide-mediated correlations can acquire a spatially organized local boundary--bulk structure. 
The interparticle distance controls where the dominant local connected correlation is formed, while the squeezing parameter controls the buildup of the connected correlations. 

\section{Steady-State Quantum Correlations} \label{sec.steady_state}

We now turn to the steady-state regime, where the transient buildup of excitations and correlations has saturated.
In this section, we examine how the connected correlations depend on the squeezing parameter $r$ and the interparticle distance $\phi$, and whether the local boundary--bulk contrast identified in the transient dynamics remains visible after relaxation.
To obtain a system-wide view of the steady-state correlations, we first consider the averages over all ordered pairs and triples of distinct emitters.
For the second-order correlation, we define
\bea
G_{\rm all}^{(2)}({\rm ss})=
\frac{1}{N(N-1)}
\sum_{i\neq j}
G_{i,j}^{(2)}({\rm ss}),
\label{Eq.G2all_ss}
\eea
while for the third-order correlation we define
\bea
G_{\rm all}^{(3)}({\rm ss})=
\frac{1}{N(N-1)(N-2)}
\sum_{\substack{i\neq j\neq k}}
G_{i,j,k}^{(3)}({\rm ss}).
\label{Eq.G3all_ss}
\eea
Here the sums include all ordered index combinations with distinct emitters, so mirror-related elements and different permutations of the same set of emitters are included.
These quantities provide system-averaged measures of the connected correlations generated across the array.
A positive value indicates that correlated excitation fluctuations dominate on average, whereas a negative value indicates that anticorrelated fluctuations dominate on average within the cumulant measure \cite{Kubo1962}.

\begin{figure}[t]
	\centering
	\includegraphics[width=0.5\textwidth]{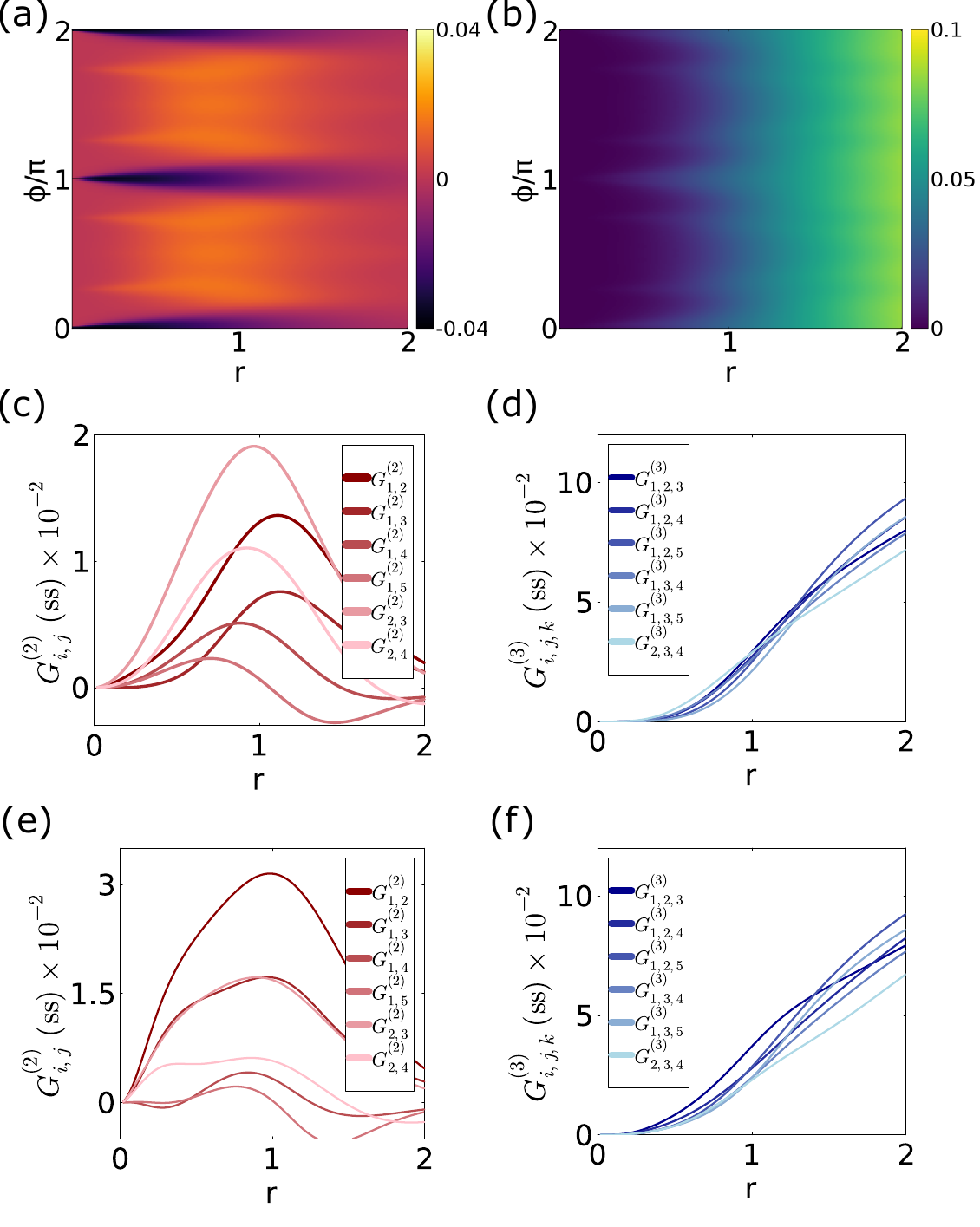}
	\caption{
		Steady-state connected correlations for $N=5$.
		(a) and (b) show heatmaps of the system-averaged second- and third-order steady-state correlations, $G_{\mathrm{all}}^{(2)}(\mathrm{ss})$ and $G_{\mathrm{all}}^{(3)}(\mathrm{ss})$, respectively, in the $(r,\phi)$ parameter space.
		(c) and (d) display representative steady-state correlations $G^{(2)}_{i,j}(\mathrm{ss})$ and $G^{(3)}_{i,j,k}(\mathrm{ss})$ as functions of $r$ for $\phi=\pi/6$, corresponding to the bulk-dominated local regime in Fig.~\ref{Fig3}.
		(e) and (f) show the corresponding steady-state correlations for $\phi=\pi/4$, corresponding to the boundary-dominated local regime.
		The remaining parameters are the same as in Fig.~\ref{Fig3}.
	}\label{Fig5}
\end{figure}

Figures~\ref{Fig5}(a) and \ref{Fig5}(b) show the system-averaged steady-state correlations in the $(r,\phi)$ parameter space for $N=5$.
The averaged second-order correlation, $G_{\rm all}^{(2)}({\rm ss})$, exhibits a pronounced dependence on $\phi$ and contains both positive and negative regions.
This sign change shows that the nonlinear reservoir can favor either correlated or anticorrelated two-emitter excitation fluctuations, depending on the interference condition set by the interparticle distance.
The largest positive values occur at intermediate squeezing strengths, whereas the correlations are suppressed for small $r$ and near selected values of $\phi$, such as around $\phi\simeq \pi$.
By contrast, $G_{\rm all}^{(3)}({\rm ss})$ remains positive in the parameter region shown and grows strongly with increasing $r$.
This behavior indicates that the second- and third-order connected correlations acquire qualitatively different steady-state signatures: the second-order correlation is more sensitive to sign-changing redistribution among pair correlations, whereas the third-order correlation is generated more directly by increasing squeezing.
This distinction is consistent with the fact that the nonlinear waveguide introduces squeezing-assisted pairing processes and distance-dependent gain, which can redistribute different orders of connected excitation statistics in different ways \cite{Karnieli2025}.

To connect the steady-state landscape with the transient local boundary--bulk contrast, we next examine representative correlation elements at fixed interparticle distances.
Figures~\ref{Fig5}(c) and \ref{Fig5}(d) show the steady-state second- and third-order correlations as functions of $r$ for $\phi=\pi/6$, corresponding to the bulk-dominated local regime identified in Fig.~\ref{Fig3}.
The representative second-order correlations in Fig.~\ref{Fig5}(c) display a nonmonotonic dependence on $r$: several pair correlations grow from zero, reach a maximum at intermediate squeezing, and then decrease as $r$ becomes larger.
The peak positions and magnitudes depend on the selected pair, showing that the steady-state pair correlations are not determined solely by emitter separation.
Instead, they reflect the combined effect of propagation-induced interference, accumulated squeezing gain, and the spatial location of the pair within the array.
This behavior extends the range-dependent correlation structure known in linear atom--nanophotonic systems to the present nonlinear setting, where the accumulated gain provides an additional control knob \cite{Jen2022,Handayana2025}.
In contrast, the representative third-order correlations in Fig.~\ref{Fig5}(d) increase more monotonically with $r$ and remain positive in the range shown, indicating a stronger direct buildup of higher-order connected correlations under increasing squeezing.

Figures~\ref{Fig5}(e) and \ref{Fig5}(f) show the corresponding steady-state correlations for $\phi=\pi/4$, corresponding to the boundary-dominated local regime in Fig.~\ref{Fig3}.
The second-order correlations in Fig.~\ref{Fig5}(e) again show nonmonotonic behavior as a function of $r$, but the dominant pair correlations differ from those in Fig.~\ref{Fig5}(c).
This confirms that the interparticle distance can redistribute the steady-state pair correlations among different correlation elements, consistent with the transient behavior in Fig.~\ref{Fig3}.
The third-order correlations in Fig.~\ref{Fig5}(f) behave similarly to those in Fig.~\ref{Fig5}(d): they remain positive and grow strongly with increasing $r$.
Thus, while second-order steady-state correlations show sign-changing and pair-dependent optimization, third-order connected correlations are generated more robustly by the squeezing-induced many-emitter excitation buildup.

Overall, Fig.~\ref{Fig5} shows that the local boundary--bulk contrast identified in the transient dynamics leaves a clear signature in the steady state.
The second-order correlations are shaped by a competition between interference, accumulated gain, and redistribution among different pair correlations.
The third-order correlations, on the other hand, show a more direct growth with the squeezing parameter.
This distinction between second- and third-order steady-state behavior highlights the role of nonlinear waveguide-mediated processes in organizing connected correlations across the emitter array.

\section{Odd--Even Comparison of Local Boundary--Bulk Contrast}
\label{sec.finite_size}

We finally examine how the steady-state local boundary--bulk contrast depends on the parity of the emitter number.
Here we focus on the second-order correlation because the position of the representative bulk nearest-neighbor pair changes between odd and even arrays. 
For $G^{(2)}$, the even array has a central nearest-neighbor bond, whereas the odd array has two mirror-related bonds adjacent to the central emitter. 
The third-order correlations also contain spatial boundary--bulk information, as shown in Secs.~\ref{sec.many_emitters} and \ref{sec.steady_state}; however, their representative bulk element follows the opposite parity structure, being centered on the middle emitter for odd $N$ and displaced from the center for even $N$. 
In this section, we use $G^{(2)}$ as the primary observable for the odd--even comparison.

\begin{figure}[b]
	\centering
	\includegraphics[width=0.5\textwidth]{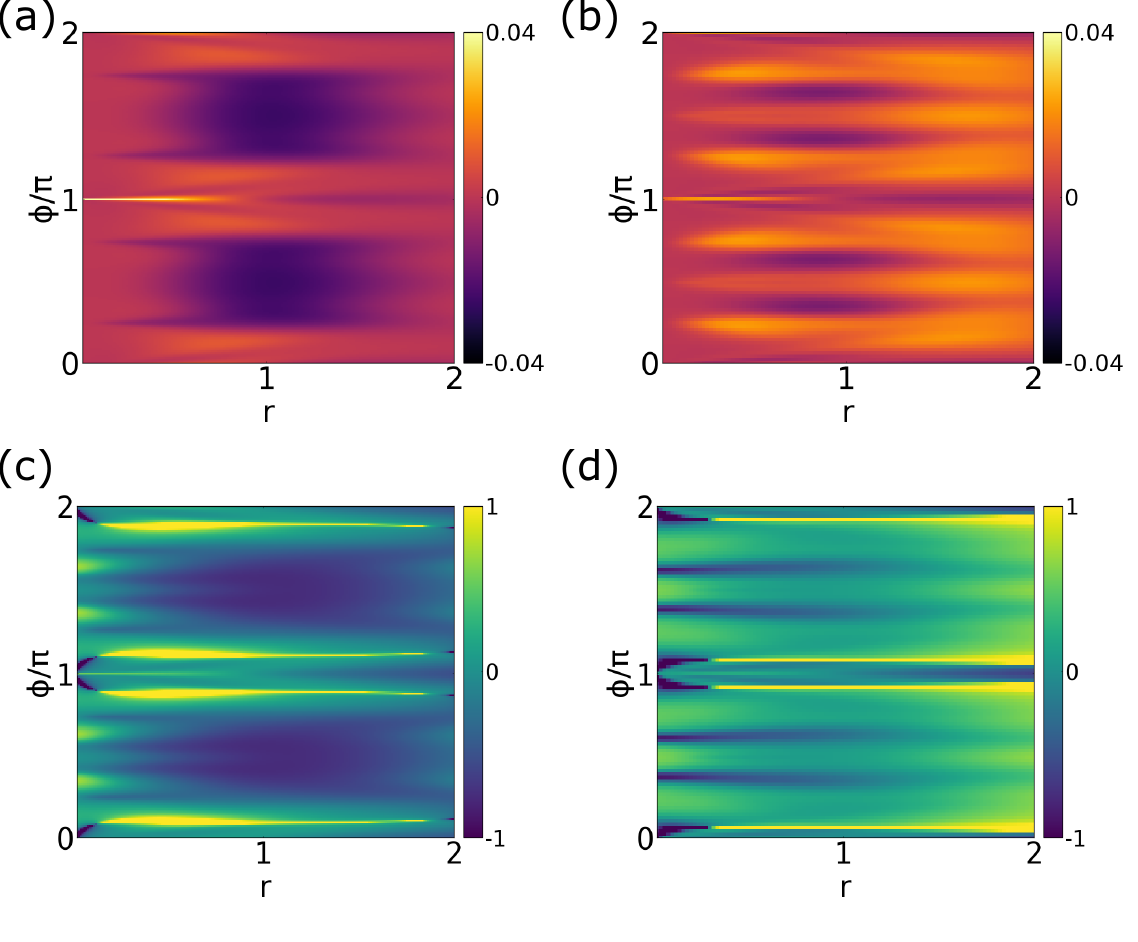}
	\caption{
		Odd--even comparison of the steady-state local boundary--bulk contrast.
		(a) and (b) show the heatmaps of the signed contrast
		$\Delta G^{(2)}_{\rm ss}
		=
		G^{(2)}_{\rm bk}({\rm ss})
		-
		G^{(2)}_{\rm bd}({\rm ss})$
		for (a) $N=5$ and (b) $N=6$.
		(c) and (d) show the corresponding normalized contrast $\eta^{(2)}_{\rm ss}$ for (c) $N=5$ and (d) $N=6$.
		Positive values indicate bulk-dominated local nearest-neighbor correlations, whereas negative values indicate boundary-dominated local nearest-neighbor correlations.
		For $N=5$, the representative bulk pair lies adjacent to the central emitter, while for $N=6$ it lies on the central bond.
		The remaining parameters are the same as in Fig.~\ref{Fig5}.
	}\label{Fig6}
\end{figure}

Following the local contrast introduced in Sec.~\ref{sec.many_emitters}, we have
\bea
\Delta G^{(2)}_{\rm ss}
=
G^{(2)}_{\rm bk}({\rm ss})
-
G^{(2)}_{\rm bd}({\rm ss}).
\label{Eq.DeltaG2_ss}
\eea
Here $G^{(2)}_{\rm bd}=G^{(2)}_{1,2}$ denotes the representative near-boundary nearest-neighbor correlation.
For $N=5$, we choose $G^{(2)}_{\rm bk}=G^{(2)}_{2,3}$, which is mirror-symmetric to $G^{(2)}_{3,4}$, whereas for $N=6$ the bulk correlation is chosen on the central bond, $G^{(2)}_{\rm bk}=G^{(2)}_{3,4}$.
To compare the relative dominance independently of the overall correlation magnitude, we also use the normalized contrast
\bea
\eta^{(2)}_{\rm ss}
=
\frac{
	G^{(2)}_{\rm bk}({\rm ss})
	-
	G^{(2)}_{\rm bd}({\rm ss})
}{
	\left|G^{(2)}_{\rm bk}({\rm ss})\right|
	+
	\left|G^{(2)}_{\rm bd}({\rm ss})\right|
}.
\label{Eq.EtaG2_ss}
\eea

Figures~\ref{Fig6}(a) and \ref{Fig6}(b) compare $\Delta G^{(2)}_{\rm ss}$ for $N=5$ and $N=6$, respectively.
For both array sizes, the contrast exhibits alternating positive and negative regions in the $(r,\phi)$ parameter space, showing that the competition between near-boundary and bulk nearest-neighbor correlations persists in the steady state.
The sign of the contrast is controlled strongly by $\phi$, reflecting the interference condition of the waveguide-mediated exchange and pairing processes.
At the same time, the contrast changes with $r$, indicating that accumulated squeezing controls the strength of the local boundary--bulk imbalance.

Compared with the odd array in Fig.~\ref{Fig6}(a), the even array in Fig.~\ref{Fig6}(b) shows a broader region with positive $\Delta G^{(2)}_{\rm ss}$.
This indicates that the central-bond placement of the bulk pair makes the bulk nearest-neighbor correlation more competitive against the near-boundary correlation over a wider range of $(r,\phi)$.
This behavior is consistent with the more symmetric coupling of the central bond to the left- and right-propagating squeezed fields.
However, the sign of $\Delta G^{(2)}_{\rm ss}$ still depends strongly on $\phi$, indicating that the odd--even parity modifies the interference-controlled boundary--bulk competition rather than eliminating it.

Figures~\ref{Fig6}(c) and \ref{Fig6}(d) show the normalized contrast $\eta^{(2)}_{\rm ss}$.
For $N=5$, the contrast alternates between positive and negative regions as $\phi$ is varied, indicating that the relative dominance of the bulk and near-boundary pairs remains strongly controlled by the interference condition.
For $N=6$, positive regions occupy a broader part of the parameter space, showing that the bulk nearest-neighbor pair remains relatively stronger even after the contrast is normalized by the total local correlation strength.
This supports the interpretation that the central-bond geometry in the even array makes the bulk pair more competitive against the near-boundary pair, rather than merely increasing the overall magnitude of the correlations.

\section{Discussion and Conclusion}\label{sec.discuss}
The results presented here extend the common framework of waveguide-mediated correlations from a passive linear reservoir to an active nonlinear reservoir.
In conventional wQED, collective decay, coherent exchange, subradiance, and photon--photon correlations arise from interference in the guided continuum \cite{Sheremet2023,Masson2020,Mahmoodian2020}.
When the reservoir is squeezed, the emitter dynamics becomes sensitive not only to emitter separations but also to the spatial phase and position of the squeezing source, as shown in squeezed-vacuum wQED \cite{You2018}.
Nonlinear wQED further changes this situation because the guided field can acquire parametric gain during propagation, producing squeezing-induced pairing terms and distance-dependent many-body interactions \cite{Karnieli2025}.
Recent work has further shown that nonlinear waveguide environments can expose otherwise hidden dark-state manifolds through output photon correlations \cite{Nadel2026}.
Our results reveal a complementary capability of the same nonlinear reservoir: accumulated squeezing generates connected excitation correlations and enables their spatial control between local near-boundary and bulk regions.
Related parametrically amplified open light--matter models also demonstrate that coherent and dissipative two-photon processes can strongly modify steady-state phases and critical behavior \cite{Zhu2026}.

The spatial redistribution found here also clarifies how a nonlinear reservoir produces distinct order-dependent responses from correlation formation in passive atom--nanophotonic systems.
In linear wQED, spatial correlations are mainly shaped by propagation phase, collective interference, and the structure of radiative eigenmodes \cite{Sheremet2023,Masson2020,Jen2022,Handayana2025}.
In the present nonlinear setting, these interference effects remain important, but they are accompanied by distance-dependent gain and squeezing-induced pairing processes.
This additional nonlinear ingredient allows the dominant local correlation to move between near-boundary and bulk regions as the interparticle distance is tuned.
The steady-state behavior further shows that this redistribution is order dependent: second-order correlations exhibit sign-changing and nonmonotonic patterns, while third-order connected correlations grow more directly with the squeezing parameter.
These results show that nonlinear wQED can not only enhance correlation strength, but also control the spatial distribution of local correlations and manipulate higher-order connected correlations.

From an experimental perspective, the predicted boundary–bulk switching and the distinct squeezing dependence of $G^{(2)}$ and $G^{(3)}$ provide concrete targets for future nonlinear wQED experiments. The required ingredients have been developed in separate contexts: strong emitter–waveguide coupling has been demonstrated in nanofiber, photonic-crystal, quantum-dot, superconducting-circuit, color-center, and integrated atom–nanophotonic platforms \cite{Morrissey2009,Vetsch2010,Solano2017,Goban2015,Douglas2015,Luxmoore2013,Arcari2014,Yalla2014,Sollner2015,Roushan2017,Wang2019,Riedel2026,Chang2018}, while parametrically driven $\chi^{(2)}$ and integrated nonlinear photonic systems provide routes to squeezed guided fields and gain engineering \cite{Helt2020,Quesada2020,Dutt2024,Weiss2026}. A direct implementation would require combining these ingredients with control over the squeezing strength, squeezing phase, and interparticle phase. The connected excitation correlations considered here would be most directly accessed through site-resolved emitter readout, while guided-output photon correlations may provide an indirect probe. Thus, the predicted local boundary–bulk contrast offers a concrete signature for testing how nonlinear gain accumulation reshapes spatial correlations in open quantum systems.

Beyond experimental verification, our results establish quantum correlations as a valuable resource for waveguide-mediated quantum information processing. The ability to selectively localize, redistribute, and enhance higher-order connected correlations through engineered squeezing provides a new degree of freedom for manipulating multipartite quantum resources, with potential applications in entanglement generation \cite{Goswami2026}, quantum state transfer, and measurement-based quantum information protocols \cite{Chien2024}. Looking ahead, combining nonlinear gain engineering with individually addressable driving fields offers exciting opportunities for programmable quantum state engineering, where tailored local excitations can dynamically shape correlation landscapes, stabilize desired many-body states, and realize reconfigurable quantum functionalities in scalable waveguide QED architectures.

\section{Acknowledgments}
We acknowledge support from the National Science and Technology Council (NSTC), Taiwan, under the Grants No. 112-2112-M-001-079-MY3, NSTC-115-
2112-M-001-035-MY3 and No. NSTC-115-2119-M-001-009, and from Academia Sinica under Grant AS-CDA-113-M04. We are also grateful for support from TG 1.2 of NCTS at NTU.

\appendix
\section{Mirror symmetry} \label{App.mirrorsymetry} 
We derive the mirror-symmetry condition used in the main text and discuss its consequences for site excitations and connected correlations. 
We start from the master equation in Eq.~(\ref{Eq.master}), with the Hamiltonian decomposition in Eq.~(\ref{Eq.H_split}) and the jump operators in Eqs.~(\ref{Eq.Lright}) and (\ref{Eq.Lleft}). 
The spatial reflection about the center of the emitter array is defined as
\bea 
j \mapsto \bar j \equiv N+1-j. \notag
\eea 
This reflection is generated by a unitary operator $P$ acting on the emitter operators as 
$P\sigma_{\pm,j}P^\dagger = \sigma_{\pm,\bar j}$. 

We first consider the exchange part of the Hamiltonian. 
The term $H_{\rm ex}$ in Eq.~(\ref{Eq.Hex}) is invariant under reflection because its coupling coefficient depends only on the relative distance between emitters. 
Under $i\mapsto\bar i$ and $j\mapsto\bar j$, one has 
$|\bar j-\bar i| = |(N+1-j)-(N+1-i)| = |j-i|$. 
Thus, the coefficient $\sin(\phi|j-i|)\cosh(\Delta r|j-i|)$ is unchanged. Furthermore, $P(\sigma^\dagger_{j}\sigma_{i})P^\dagger = \sigma_{+,\bar j}\sigma_{-,\bar i}, $ and $P(\sigma^\dagger_{i}\sigma_{j})P^\dagger = \sigma_{+,\bar i}\sigma_{-,\bar j}.$ Since $\bar i$ and $\bar j$ are dummy summation indices, relabeling them back to $i$ and $j$ gives 
\bea 
P H_{\rm ex}P^\dagger = H_{\rm ex}. \notag
\eea 

The nontrivial constraint comes from the squeezing-induced pairing Hamiltonian $H_{\rm pair}$ in Eq. (\ref{Eq.Hpair}). Under reflection, the center-of-mass phase transforms according to $i+j \mapsto \bar i+\bar j = 2(N+1)-(i+j).$ Therefore, 
\bea 
e^{i\phi(i+j)} \mapsto e^{i2(N+1)\phi}e^{-i\phi(i+j)}, 
\eea 
and 
\bea 
e^{-i\phi(i+j)} \mapsto e^{-i2(N+1)\phi}e^{i\phi(i+j)}. 
\eea 
At the same time, the two propagation channels are exchanged, 
\bea 
\Theta(j-i) \mapsto \Theta(i-j), \qquad \Theta(i-j) \mapsto \Theta(j-i). \notag
\eea 
Thus, the reflection maps the right-propagating pairing channel into the left-propagating one, and vice versa. To obtain the phase condition explicitly, consider the right-propagating pair-creation term in Eq. (\ref{Eq.Hpair}), 
\bea \Theta(j-i) e^{-i\theta_{\rightarrow}} e^{i\phi(i+j)} \sigma^\dagger_{j}\sigma^\dagger_{i}. 
\eea 
After applying $P$ and relabeling the reflected indices back to $(i,j)$, this term becomes 
\bea \Theta(i-j) e^{-i\theta_{\rightarrow}} e^{i2(N+1)\phi} e^{-i\phi(i+j)} \sigma^\dagger_{j}\sigma^\dagger_{i}. 
\eea 

For this reflected term to have the same form as the corresponding left-propagating pair-creation term, we require $e^{-i\theta_{\rightarrow}} e^{i2(N+1)\phi} = e^{-i\theta_{\leftarrow}}.$ Equivalently, 
\bea 
\theta_{\rightarrow}-\theta_{\leftarrow} = 2(N+1)\phi+2m\pi, \qquad m\in\mathbb{Z}. \label{Eq.mirror_condition_appendix} 
\eea 
The pair-annihilation terms give the same condition. Therefore, 
\bea 
P H P^\dagger = H \notag
\eea 
when Eq. (\ref{Eq.mirror_condition_appendix}) is satisfied. 

We next show that the same phase condition also preserves the dissipative part of the master equation. From Eq. (\ref{Eq.Lright}), the right-propagating jump operator can be written as 
\bea 
L^\rightarrow = \sqrt{\gamma} \sum_{j=1}^{N} \left[ e^{-i\phi j}C_j\sigma_{j} - i e^{i\phi j}e^{-i\theta_{\rightarrow}} S_j\sigma^\dagger_{j} \right], 
\eea 
where $C_j = \cosh[\bar r+\Delta r(j-1)],$ and $S_j = \sinh[\bar r+\Delta r(j-1)].$ Under reflection, these coefficients become 
\bea 
C_j \mapsto \widetilde C_j = \cosh[\bar r+\Delta r(N-j)], 
\eea 
and 
\bea 
S_j \mapsto \widetilde S_j = \sinh[\bar r+\Delta r(N-j)]. 
\eea 
Therefore, 
\bea \notag
P L^\rightarrow P^\dagger = \sqrt{\gamma} \sum_{j=1}^{N} \Big[&& e^{-i\phi(N+1-j)} \widetilde C_j\sigma_{j} \\\notag
&&-ie^{i\phi(N+1-j)} e^{-i\theta_{\rightarrow}} \widetilde S_j\sigma^\dagger_{j} \Big]. \notag
\eea 
Using $e^{-i\phi(N+1-j)} = e^{-i\phi(N+1)}e^{i\phi j},$ and $e^{i\phi(N+1-j)} = e^{i\phi(N+1)}e^{-i\phi j},$ we obtain 
\bea \nonumber
P L^\rightarrow P^\dagger =&& e^{-i\phi(N+1)} \sqrt{\gamma} \sum_{j=1}^{N} \Big[ e^{i\phi j}\widetilde C_j\sigma_{j} \\\nonumber
&&- i e^{-i\phi j} e^{-i[\theta_{\rightarrow}-2(N+1)\phi]} \widetilde S_j\sigma^\dagger_{j} \Big]. \nonumber
\eea 
When Eq. (\ref{Eq.mirror_condition_appendix}) is imposed, this reduces to 
\bea 
P L^\rightarrow P^\dagger = e^{-i\phi(N+1)} L^\leftarrow. \eea Similarly, one obtains \bea P L^\leftarrow P^\dagger = e^{i\phi(N+1)} L^\rightarrow. 
\eea 
The remaining factors $e^{\pm i\phi(N+1)}$ are global phases of the jump operators. They do not change the Lindblad dissipator because $\mathcal{D}_{e^{i\chi}L}[\rho] = \mathcal{D}_{L}[\rho].$ 

Consequently, the full Liouvillian is invariant under mirror reflection when the phase condition in Eq. (\ref{Eq.mirror_condition_appendix}) is satisfied. This Liouvillian symmetry has direct consequences for physical observables. If the initial state is mirror symmetric, $P\rho(0)P^\dagger = \rho(0),$ then the density matrix remains mirror symmetric at all times, $P\rho(t)P^\dagger = \rho(t).$ This condition is satisfied, for example, by the ground-state initial condition used in this work. For the site excitation, one then obtains 
\bea 
\langle n_j(t)\rangle = {\rm Tr}[\rho(t)n_j]. 
\eea 
Using $P n_j P^\dagger=n_{\bar j}$ and $P\rho(t)P^\dagger=\rho(t)$, we find 
\bea 
\langle n_j(t)\rangle = \langle n_{\bar j}(t)\rangle. 
\eea 
Thus, mirror-related emitters have identical excitation dynamics. The same argument applies to the density-density moments entering the connected correlations. 
Since $P(n_i n_j)P^\dagger = n_{\bar i}n_{\bar j}$, we have
\bea 
\langle n_i n_j\rangle = \langle n_{\bar i}n_{\bar j}\rangle.
\eea 
Together with the mirror symmetry of the single-site populations, this gives
\bea 
G^{(2)}_{i,j}(t) = G^{(2)}_{\bar i,\bar j}(t).
\eea
For third-order connected correlations, the reflection gives 
$P(n_i n_j n_k)P^\dagger = n_{\bar i}n_{\bar j}n_{\bar k}$. 
All lower-order moments entering the third-order cumulant transform in the same way, and hence
\bea 
G^{(3)}_{i,j,k}(t) = G^{(3)}_{\bar i,\bar j,\bar k}(t).
\eea
These relations justify grouping mirror-related correlation elements together. 
For example, the near-boundary nearest-neighbor pair $(1,2)$ is equivalent to $(N-1,N)$, while the bulk nearest-neighbor pair $(2,3)$ is equivalent to $(N-2,N-1)$ under the mirror-symmetric dynamics. 
This reduction is used in the main text to select representative local correlations near the boundary and in the bulk.

For the equal-phase choice $\theta_{\rightarrow}=\theta_{\leftarrow}$, the phase condition in Eq. (\ref{Eq.mirror_condition_appendix}) becomes 
\bea 
2(N+1)\phi = 2m\pi, \qquad m\in\mathbb{Z}. \notag
\eea 
Therefore, 
\bea 
\phi = \frac{m\pi}{N+1}, \qquad m\in\mathbb{Z}. 
\eea 
The wavelength-spaced case $\phi=2\pi$ is one particular member of this family, because it automatically satisfies the equal-phase condition for any integer $N$.

\end{document}